\documentclass[twocolumn]{aastex631}

\begin{document}

\title{ADF22-WEB: Detection of a molecular gas reservoir in a massive quiescent galaxy located in a $z\approx3$ proto-cluster core}

\author[0000-0003-1937-0573]{Hideki Umehata}
\affiliation{Institute for Advanced Research, Nagoya University, Furocho, Chikusa, Nagoya 464-8602, Japan}
\affiliation{Department of Physics, Graduate School of Science, Nagoya University, Furocho, Chikusa, Nagoya 464-8602, Japan}
\author{Mariko Kubo}
\affiliation{Astronomical Institute, Tohoku University, 6-3, Aramaki, Aoba, Sendai, Miyagi, 980-8578, Japan}
\author{Kouichiro~Nakanishi}
\affiliation{National Astronomical Observatory of Japan, 2-21-1 Osawa, Mitaka, Tokyo 181-8588, Japan}
\affiliation{Department of Astronomical Science, The Graduate University for Advanced Studies, SOKENDAI, 2-21-1 Osawa, Mitaka, Tokyo
181-8588, Japan}



\begin{abstract}
We present a study of the molecular gas reservoirs and dust contents in three quiescent galaxies (QGs) located in the core of the $z=3.09$ SSA22 proto-cluster. Using the Atacama Large Millimeter/submillimeter Array (ALMA), we detect CO(3--2) emission in one galaxy, ADF22-QG1, marking the first direct detection of molecular gas in a quiescent galaxy from the early universe. 
The detected galaxy, ADF22-QG1, has a molecular gas mass of log$M_{\rm H_2}$/M$_\odot = 10.26 \pm 0.07$ assuming a CO-to-H$2$ conversion factor $\alpha_{\rm CO} = 4.4$ (log$M_{\rm H_2}$/M$_\odot = 9.52 \pm 0.07$ for $\alpha_{\rm CO} = 0.8$), corresponding to a gas mass fraction of $f_{\rm gas} \approx 14\%$ (2.5\%). 
The gas-to-dust ratio $\delta _{\rm gdr}\gtrsim170$ 
 ($\delta_{\rm gdr}\gtrsim30$) for $\alpha_{\rm CO} = 4.4$ ($\alpha_{\rm CO} =0.8$) is also derived for the first time for a QG at the epoch.
For the other two galaxies, ADF22-QG2 and ADF22-QG3, non detections of CO(3--2) emission provide upper limits, $f_{\rm gas} \approx 17\%$ (3.1\%) and $f_{\rm gas} \approx 13\%$ (2.4\%), respectively.
The inferred gas-consumption history of ADF22-QG1, based on its star-formation history, suggests that (i) dusty star-forming galaxies (DSFGs) at $z = 4$--$6$ are plausible progenitors, and (ii) the cessation of gas accretion from cosmic web filaments plays an important role in their evolution to quenched systems. Furthermore, the presence of a detectable molecular gas reservoir in ADF22-QG1 indicates that additional mechanisms, such as morphological quenching, may be required to fully explain its quiescent nature.
\end{abstract}




\section{Introduction} \label{sec:intro}

To illustrate the formation and evolution of massive galaxies in the early universe is a key goal in modern astronomy. Over the past decade, spectroscopic confirmations of massive quiescent galaxies (QGs) at $z=3-5$ with $\log(M_*/M_\odot) \approx 10-11$ have been reported (\citealt{2012ApJ...759L..44G}; \citealt{2017Natur.544...71G}; \citealt{2018A&A...611A..22S}; \citealt{2018A&A...618A..85S}; \citealt{2020ApJ...890L...1F}; \citealt{2020ApJ...889...93V}; \citealt{2021ApJ...919....6K}; \citealt{2022ApJ...935...89K}; \citealt{2023Natur.619..716C}; \citealt{2024ApJ...974..145S}; \citealt{2024MNRAS.534..325C}; \citealt{2024NatAs.tmp..284D}; \citealt{2024MNRAS.534.3552S}; \citealt{2024NatAs...8.1443D}). These findings reveal that mass assembly and quenching processes occur at significantly earlier times in cosmic history. Analyses of photometry and spectra suggest a history of active star formation followed by a quenching event (e.g., \citealt{2021ApJ...919....6K}). While dusty star-forming galaxies (DSFGs; for a review, see \citealt{2014PhR...541...45C}) with various levels of dust continuum detections are proposed as potential progenitors (e.g., \citealt{2020ApJ...889...93V}), the precise evolutionary pathways and quenching mechanisms remain unclear.

One critical missing piece of this puzzle is the molecular gas reservoir in QGs. While detections of CO lines have been reported for some QGs at $z\sim1$ (\citealt{2018ApJ...860..103S}; \citealt{2021ApJ...908...54W}; \citealt{2022ApJ...925..153B}), reliable detections of molecular gas tracers at higher redshifts remain elusive (e.g., \citealt{2023A&A...678L...9D}; \citealt{2024arXiv240519401S}). Some studies have attempted to constrain molecular gas masses based on the (non-)detection of dust continuum (e.g., \citealt{2021Natur.597..485W}), but these efforts are hindered by uncertainties in the conversion factors between dust mass and molecular gas mass (\citealt{2019MNRAS.490.1425L}; \citealt{2021ApJ...922L..30W}).

\begin{figure*}
\epsscale{0.8}
\plotone{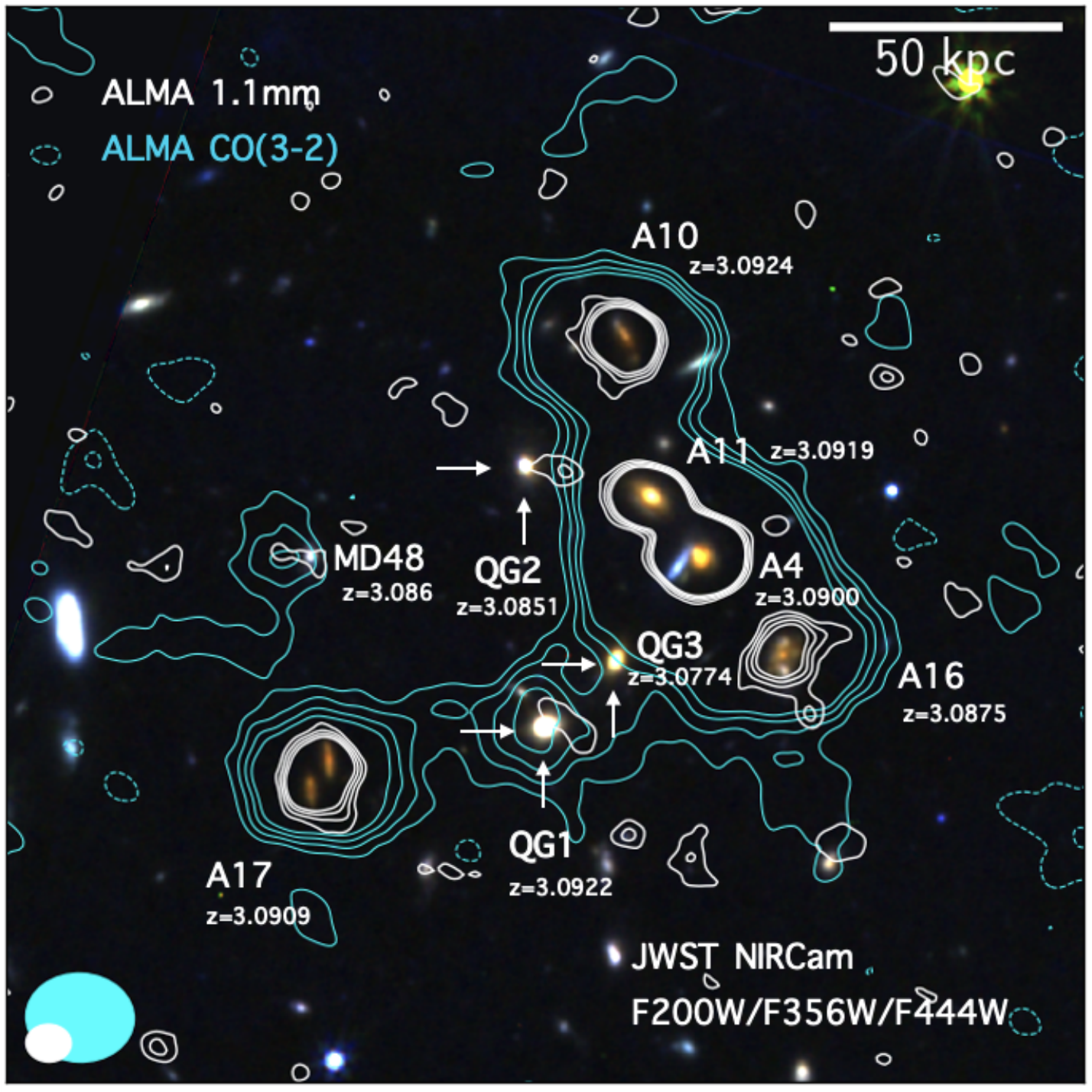}
\caption{
NIRCam color image (F200W/F356W/F444W) of the AzTEC14 dense galaxy group at $z\approx3.09$ in ADF22 (\citealt{2014MNRAS.440.3462U}; \citealt{2015ApJ...799...38K}). Cyan (white) contours indicate CO(3--2) and 1.1\,mm emission at levels of 2$\sigma$, 3$\sigma$, ..., $5\sigma$. Confirmed proto-cluster members, including DSFGs (ADF22.A4, ADF22.A10, ADF22.A11, ADF22.A16, ADF22.A17 (\citealt{2017ApJ...835...98U}; \citeyear{2019Sci...366...97U})), an LBG (MD48 (\citealt{1998ApJ...492..428S})), and QGs (ADF22-QG1, ADF22-QG2, ADF22-QG3 (\citealt{2021ApJ...919....6K}; \citeyear{2022ApJ...935...89K})) are labeled with their $z_{\rm spec}$ values. ADF22-QG1 is detected in CO(3--2) but not in the dust continuum. Neither the dust continuum nor the CO(3--2) line is significantly detected in ADF22-QG2 or ADF22-QG3.
}
\label{fig:alma_map}
\end{figure*}

In this Letter, we present observations of three QGs at $z\sim3$ and the first detection of a molecular gas reservoir associated with a QG. The targets are located in ADF22 (``ALMA Deep Field in SSA22'', \citealt{2015ApJ...815L...8U}; \citeyear{2017ApJ...835...98U}; \citeyear{2018PASJ...70...65U}), the core region of the SSA22 protocluster at $z=3.09$ (\citealt{1998ApJ...492..428S}; \citealt{2004AJ....128.2073H}; \citealt{2005ApJ...634L.125M}). This region contains a dense galaxy group at $z\approx3.09$, referred to as the AzTEC14 group, originally identified as an AzTEC/ASTE source (\citealt{2009Natur.459...61T}; \citealt{2014MNRAS.440.3462U}), located along Mpc-scale cosmic web gas filaments traced by Ly\,$\alpha$ in emission (\citealt{2019Sci...366...97U}). The AzTEC14 group hosts a variety of galaxies, including an LBG (\citealt{1998ApJ...492..428S}), DSFGs (\citealt{2015ApJ...815L...8U}; \citeyear{2017ApJ...835...98U}), and three QGs, ADF22-QG1, ADF22-QG2, and ADF22-QG3 (\citealt{2021ApJ...919....6K}; \citeyear{2022ApJ...935...89K}). One DSFG (ADF22.A4) and one QG (ADF22-QG2) have X-ray AGNs (\citealt{2009ApJ...691..687L}; \citeyear{2009MNRAS.400..299L}).

We adopt a standard concordance cosmology with $H_0=70$\,km\,s$^{-1}$\,Mpc$^{-1}$, $\Omega_{\rm m}=0.30$, and $\Omega_\Lambda=0.70$.

\section{Observation and Reduction} \label{sec:obs}

We observed the CO(3--2) line ($\nu_{\rm rest} = 345.796\,\mathrm{GHz}$) using ALMA toward the AzTEC14 group. The observations were conducted as part of ALMA project 2023.1.01206.S (PI: Umehata). The phase center was set to ($\alpha$, $\delta$) = (22h17m37.0s, 0d18m20.7s), corresponding to the position of a bright DSFG, ADF22.A4. The observations were performed over 21 runs during ALMA Cycle 10 (between January and September 2024) using the Band-3 receiver. A range of array configurations (C-1 to C-5) was employed. The total on-source integration time was 17 hours. 
The correlator was configured with two spectral windows of 1.875\,GHz bandwidth (dual polarization) per sideband, separated by 8\,GHz. Each spectral window was centered at 85.03, 86.86, 97.03, and 98.86\,GHz, respectively. Data reduction was performed using version 6.5.4 of the CASA software package (\citealt{2007ASPC..376..127M}; \citealt{2022PASP..134k4501C}). Imaging was carried out using the CLEAN algorithm via the \texttt{tclean} task with natural weighting, down to $2\sigma$ with auto-masking. The resulting synthesized beam size was $3.00^{\prime\prime} \times 2.46^{\prime\prime}$ (PA = 87\,deg). The typical rms noise level at the phase center was 33\,$\mu$Jy\,beam$^{-1}$ per 100\,km\,s$^{-1}$ channel at around 85\,GHz.

The AzTEC14 group was also covered by a deep 1.1\,mm mosaic (\citealt{2021ApJ...919....6K}; S.Huang et al. in preparation) and observed with JWST/NIRCam (F115W, F200W, F356W, and F444W) as a part of the ``ADF22-WEB'' (\citealt{2024arXiv241022155U}; \citeyear{2025arXiv250201868U}). Fig.~\ref{fig:alma_map} shows the NIRCam color image and 1.1\,mm contours, exhibit the distributions of group member galaxies.

\section{Results} \label{sec:results}

\begin{figure}
\epsscale{1}
\plotone{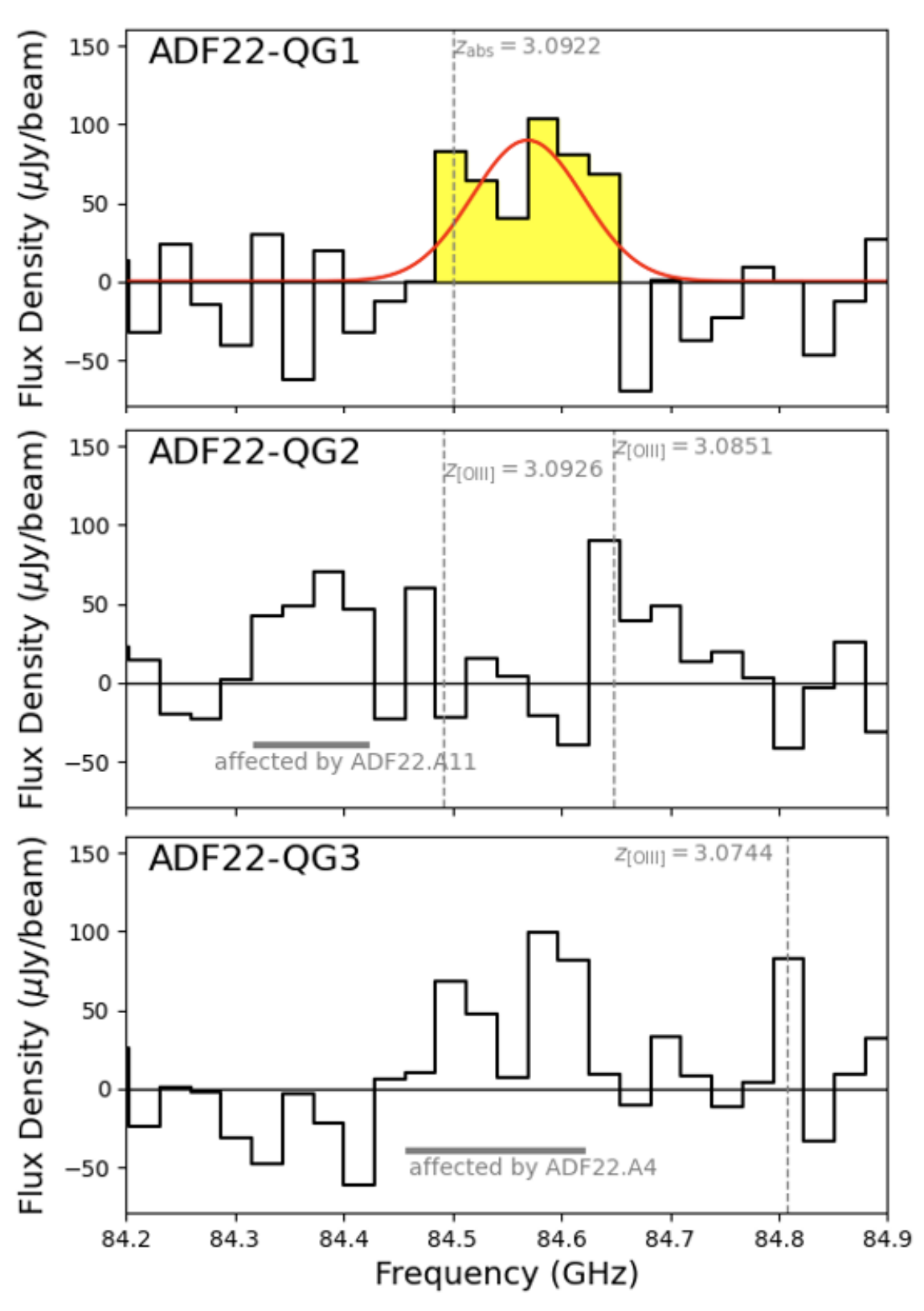}
\caption{
CO(3--2) spectra of the three QGs in the AzTEC14 group. ADF22-QG1 is detected in CO(3--2), with a redshift broadly consistent with that determined from Balmer absorption lines (\citealt{2021ApJ...919....6K}). The best-fit Gaussian profile is shown as a red solid line. The yellow-shaded region indicates the frequency range used to create the moment-0 map. No significant CO(3--2) detection is found for ADF22-QG2 or ADF22-QG3.
}
\label{fig:alma_spectra}
\end{figure}

\begin{figure*}
\epsscale{1.16}
\plotone{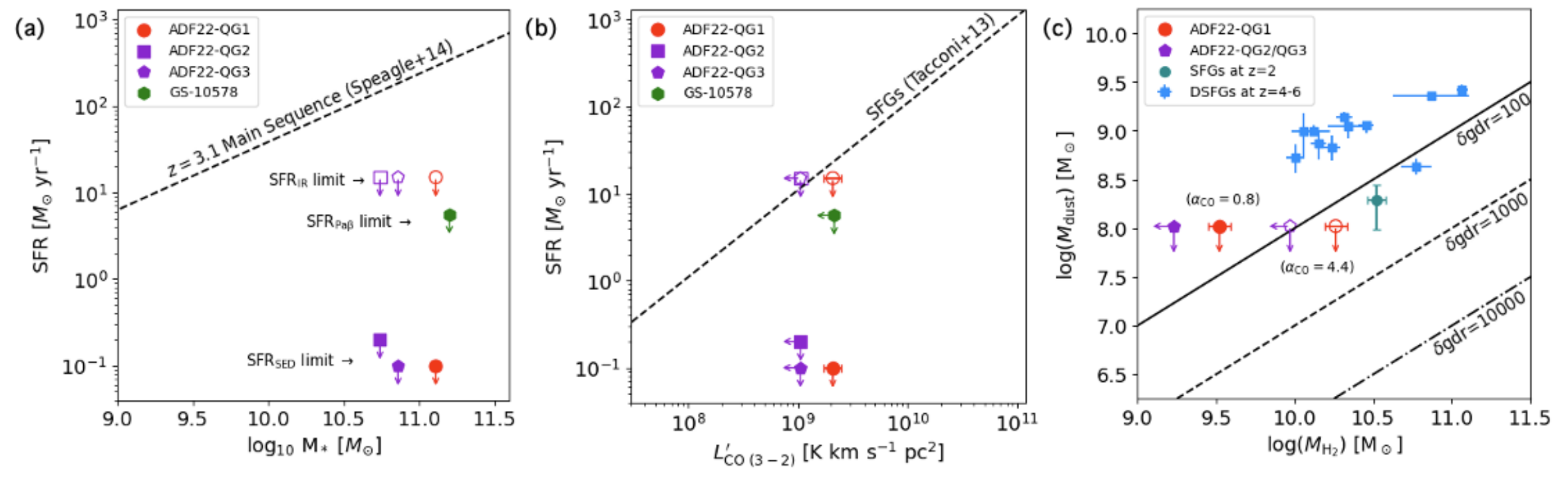}
\caption{
(a) The relation between stellar mass and SFR for the three ADF22 QGs at $z=3.09$. Their SEDs indicate that they are nearly completely quenched, in sharp contrast to the star-forming main sequence (\citealt{2014ApJS..214...15S}). The non-detections in the ALMA 1\,mm map provide independent constraints. For comparison, GS-10578, a quiescent system at $z=3.064$, is also shown (\citealt{2024arXiv240519401S}).
(b) The relation between CO(3--2) luminosity and SFR. The SED-based SFR suggests that star formation is significantly lower than expected based on the molecular gas mass, assuming the typical scaling for star-forming galaxies (\citealt{2013ApJ...768...74T}).
(c) The relation between molecular gas mass and dust mass for ADF22-QGs, shown for two assumptions of $\alpha_{\rm CO}$. For comparison, dusty star-forming galaxies (DSFGs) at $z=4-6$ from the literature (\citealt{2020ApJ...895...81R}; \citeyear{2021ApJ...907...62R}; \citealt{2020A&A...640L...8U}; \citealt{2021MNRAS.501.3926B}; \citealt{2024ApJ...961..226L}; \citealt{2024ApJ...961...69S}) and normal SFGs at $z=2$ (\citealt{2023A&A...670A.138P}) are plotted. The current depth of the dust continuum emission may not be sufficient to put a tight constraint on $\delta$gdr.
}
\label{fig:gas_dust}
\end{figure*}

The CO (3--2) emission line associated with the three QGs was searched for using the deep Band-3 cube, resulting in a detection in the most massive galaxy, ADF22-QG1. Fig~\ref{fig:alma_map} shows the CO(3--2) moment-0 map, where the flux was integrated across $\Delta v=600$\,km\,s$^{-1}$ to fully cover the emission in ADF22-QG1. Figure~\ref{fig:alma_spectra} presents the extracted spectra for the three QGs.
For ADF22-QG1, the CO(3--2) emission is detected at a significance level of 5.6$\sigma$, spatially overlapping with the stellar emission (Fig.~\ref{fig:alma_map}). A single Gaussian fit to the spectrum yields $z_{\rm CO}=3.0889\pm0.0007$, which is broadly consistent with the previously reported redshift based on Balmer absorption lines ($z_{\rm abs}=3.0922_{-0.0004}^{+0.0008}$; \citealt{2021ApJ...919....6K}). The velocity dispersion is measured as $\sigma_{\rm gas}=180\pm50$\,km\,s$^{-1}$, which is smaller than the stellar velocity dispersion $\sigma_*\approx320$\,km\,s$^{-1}$ (\citealt{2021ApJ...919....6K}). Combined with the detection of disk components in stellar emission (Kubo et al., in preparation), this may suggest that the molecular gas distribution and kinematics are not identical as those of the stellar bulge components, which may be more centrally concentrated.
For the remaining two QGs, ADF22-QG2 and ADF22-QG3, the CO(3--2) emission was not significantly detected, although very marginal signals ($\sim2\sigma$) are identified at the expected peak frequency based on the [O\,{\sc iii}]\,$\lambda 5007$ detections. 

The total line intensity of ADF22-QG1 was measured as peak flux density in the moment-0 map, which was used to estimate the molecular gas masses $M_{\rm H_2}$. In the calculation we assumed the brightness temperature ratio $r_{31}=0.5$ (\citealt{2013MNRAS.429.3047B}) to evaluate the ground state CO luminosity and the two types of CO--H$_2$ conversion factor $\alpha_{\rm CO}$ to convert the measured CO luminosity to $M_{\rm H_2}$, $\alpha_{\rm CO}=4.4$ (for a Milkey-Way like galaxies) and $\alpha_{\rm CO}=0.8$ (for a ULIRG-like galaxies). 
The derived molecular gas mass and fraction of ADF22-QG1 are $\log M_{\rm mol}/M_\odot=10.26\pm0.07$  and $\log M_{\rm mol}/M_\odot=9.52\pm0.07$ for $\alpha_{\rm CO}=4.4$ and 0.8, respectively (Table~\ref{tab:measurement}). 
To estimate upper limits for ADF22-QG2 and QG3, we created a pseudo narrow-band image for each with $\Delta v=500$\,km\,s$^{-1}$\footnote{Although ADF22-QG2 exhibits two emission line peaks as reported in \cite{2015ApJ...799...38K}, we adopt $z_{\rm [OIII]}=3.0851$, which is the most dominant peak.} and measured $3\sigma$ upper limits. Bright emission associated with DSFGs does not significantly contaminate the upper limit due to a velocity offset, as indicated in Figure~\ref{fig:alma_spectra}.
The measured upper limits are summarized in Table~\ref{tab:measurement}.

We updated the fits of ADF22 QGs with the SED models with parametric SFHs by using FAST$++$ (\citealt{2021ApJ...919....6K};  \citeyear{2022ApJ...935...89K}), incorporating new JWST photometry (Table~\ref{tab:measurement}, M. Kubo et al. in preparation for more details). 
None of the three ADF22-QGs were significantly detected in the 1.1\,mm dust continuum (Fig~\ref{fig:alma_map}). The non detections place a $3\sigma$ point-source limit of $S_{\rm 1.1mm} <75$\,$\mu$Jy (\citealt{2021ApJ...919....6K}; S. Huang et al., in preparation), which provides a lough upper limit on $SFR_{\rm IR}\sim15$\,M$_*$\,yr$^{-1}$ (\citealt{2021ApJ...919....6K}). We also estimate upper limits on the dust mass $M_{\rm d}$, assuming the dust mass absorption coefficient $\kappa_{850}=0.05$\,m$^{2}$\,kg$^{-1}$ (\citealt{2003ARA&A..41..241D}), $\beta=2.0$ , and  dust temperature of $T_{\rm d}=20$\,K (\citealt{2021A&A...647A..33M}). The derived values are summarized in Table~\ref{tab:measurement}. 

The dynamical mass of ADF22-QG1 is estimated as $M_{\rm dyn}=K_n K_q \sigma_*^2 r_e / G$ where $K_n$ and $K_q$ are scaling factor as a function of S\'ersic index $n$ and axis ratio $q$, and $\sigma_*$ is the stellar velocity dispersion and $r_e$ is the effective radius (\citealt{2022ApJ...936....9V}). The measurements of the galaxy (\citealt{2021ApJ...919....6K}; M. Kubo et al. in preparation) provides $M_{\rm dyn}\approx(2\pm0.5)\times 10^{11}$\,M$_\odot$. 
The sum of $M_*$, $M_{\rm H_2}$, and the dark matter mass $M_{\rm DM}$ (assumed to be 10--20\% of $M_{\rm dyn}$; e.g., \citealt{2017Natur.543..397G}) does not exceed $M_{\rm dyn}$ within the uncertainties for either case with $\alpha_{\rm CO}=4.4$ or 0.8.

\section{Discussion} \label{sec:dis}

\begin{figure*}
\epsscale{1.0}
\plotone{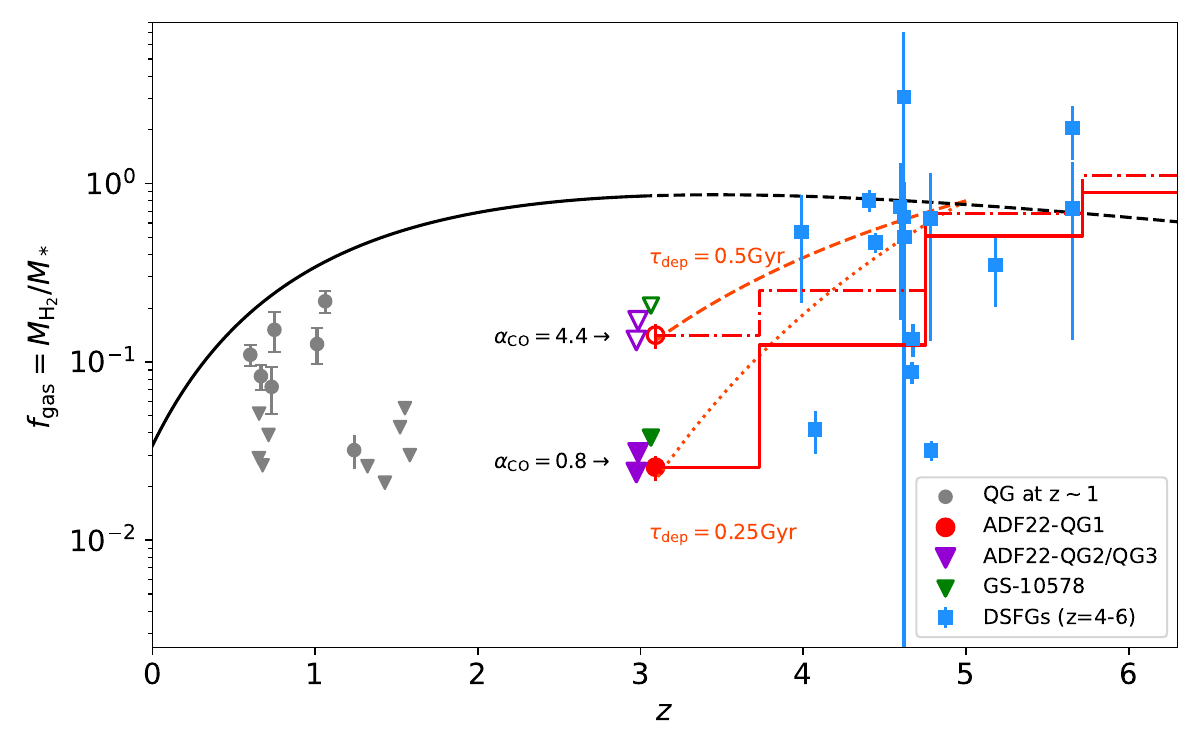}
\caption{
Gas mass fraction ($f_{\rm gas}=M_{\rm H_2}/M_*$) as a function of redshift for ADF22-QGs and other QGs at $z\gtrsim1$ (\citealt{2018ApJ...860..103S}; \citealt{2021ApJ...908...54W}; \citealt{2021ApJ...909L..11B}; \citealt{2022ApJ...940...39W} for $z\sim1$, \citealt{2024arXiv240519401S} for $z\sim3$). For QGs at $z\sim3$, filled symbols represent cases with $\alpha_{\rm CO}=4.4$, while empty symbols indicate $\alpha_{\rm CO}=0.8$. For comparison, DSFGs at $z\sim4-6$ (\citealt{2020ApJ...895...81R}; \citeyear{2021ApJ...907...62R}; \citealt{2020A&A...640L...8U}; \citealt{2021MNRAS.501.3926B}; \citealt{2024ApJ...961..226L}; \citealt{2024ApJ...961...69S}) are also plotted. The black solid and dashed lines show the scaling relation suggested by \cite{2020ARA&A..58..157T} for $z=1-3$ star-forming galaxies and its extrapolation. The red dashed (dotted) curves represent models with constant $t_{\rm dep}=0.45$\,Gyr ($t_{\rm dep}=0.25$\,Gyr), assuming external gas accretion halts at $z=5$. The red solid and dot-dashed lines indicate the evolution of gas fraction based on the star-formation history from \cite{2021ApJ...919....6K}, assuming no gas inflows, for $\alpha_{\rm CO}=4.4$ and 0.8, respectively.The gas fractions of QGs at $z\sim3$ can be explained if gas accretion onto DSFGs at $z=4-6$ is halted for some reason and the majority of the gas reservoirs is consumed afterward.
}
\label{fig:fgas}
\end{figure*}

Fig.~\ref{fig:gas_dust} presents the physical parameters derived for ADF22 QGs in comparison with other star-forming galaxies and QGs. The three ADF22 QGs are classified as quenched galaxies regardless of the SFR tracers used (Fig.~\ref{fig:gas_dust}a; see also \citealt{2021ApJ...919....6K}; \citeyear{2022ApJ...935...89K}). 
Fig.~\ref{fig:gas_dust}b indicates that ADF22-QG1 exhibits a significantly lower SFR as inferred from the CO(3--2) line luminosity ($L^{\prime}_{\rm CO(3-2)}$) than expected for a typical star-forming galaxy (e.g., \citealt{2013ApJ...768...74T}), suggesting suppressed star formation despite the presence of molecular gas. 
Fig.~\ref{fig:gas_dust} also constrains the gas-to-dust ratio, $\delta_{\rm gdr}=M_{\rm H_2}/M_{\rm d}$. As shown in Fig.~\ref{fig:gas_dust}c, we derive $\delta_{\rm gdr} \gtrsim 170$ ($\delta_{\rm gdr} \gtrsim 30$) for $\alpha_{\rm CO}=4.4$ ($\alpha_{\rm CO}=0.8$) in ADF22-QG1. Models predict that QGs at high redshift may have significantly higher values ($\delta_{\rm gdr} \gtrsim 500$--$1000$) compared to star-forming galaxies ($\delta_{\rm gdr} \sim 50$--$100$) (\citealt{2021ApJ...922L..30W}; see also \citealt{2019MNRAS.490.1425L}). This discrepancy could arise from the rapid destruction of dust in quiescent galaxies (e.g., via thermal sputtering and shocks from supernovae), which occurs more quickly than the depletion of molecular gas (\citealt{2021ApJ...922L..30W}). 
The current constraint, which has been obtained for a QG at high redshift for the first time, may not be sufficient, and a more sensitive dust continuum survey will be crucial for critically testing this prediction.

Fig~\ref{fig:fgas} summarizes the gas mass fraction ($f_{\rm gas}=M_{\rm H_2}/M_*$) of QGs at $z\gtrsim1$ from the literature (\citealt{2018ApJ...860..103S}; \citealt{2021ApJ...908...54W}; \citealt{2021ApJ...909L..11B}; \citealt{2022ApJ...940...39W}; \citealt{2024arXiv240519401S}). 
The measurements for GS-10578 at $z=3.064$ (\citealt{2024arXiv240519401S}) were rescaled with $\alpha_{\rm CO}=4.4$ (and $\alpha_{\rm CO}=0.8$), $r_{31}=0.5$, and $\Delta v = 500$\,km\,s$^{-1}$ for consistency with the ADF22-QGs and other $z\sim1$ QGs. 
As shown, the CO detection from ADF22-QG1 at $z=3.09$ provides a unique and invaluable opportunity to investigate the formation processes of these massive, passive populations in the early universe at $z\gtrsim3$.

In Fig~\ref{fig:fgas}, we present two toy models of the evolution of $f_{\rm gas}$ in a closed box system (i.e., no net inflow or outflow), assuming no additional external gas supply exceeding the amount of outflow (e.g., from the cosmic web). The first model assumes a constant depletion timescale (\citealt{2021ApJ...908...54W}). The dashed red curves in Fig~\ref{fig:fgas} show the model tracks with constant $t_{\rm dep}=0.5$ and 0.25\,Gyr. The second model, shown with solid and dot-dashed lines, is derived from the star-formation history obtained in a non-parametric manner by \cite{2021ApJ...919....6K} \footnote{We scaled the star-formation rate to match the stellar mass estimate with that of FAST++.}, simply assuming that molecular gas mass was converted to stellar mass at a rate identical with 
star-formation rate (\citealt{2024arXiv240519401S}).
In Fig~\ref{fig:fgas}, we also show the scaling relation of $f_{\rm gas}$ derived by \cite{2020ARA&A..58..157T} for star-forming galaxies with log$M_*$/M$_\odot=11$ at $z=1-3$. Additionally, we present the observed values of $f_{\rm gas}$ for DSFGs at $z\sim4-6$ (\citealt{2020ApJ...895...81R}; \citeyear{2021ApJ...907...62R}; \citealt{2020A&A...640L...8U}; \citealt{2021MNRAS.501.3926B}; \citealt{2024ApJ...961..226L}; \citealt{2024ApJ...961...69S}), with gas masses rescaled using $\alpha_{\rm CO}=1$. 
As shown, both types of model tracks predict progenitors with $f_{\rm gas}$ broadly consistent with the observed DSFGs and the extrapolated scaling relations. This result suggests that (some of) DSFGs at $z=4-6$ are progenitors of massive QGs at $z\sim3$, in line with recent works based on different evidence, such as star-formation histories and cosmic volume densities (e.g., \citealt{2020ApJ...889...93V}; \citealt{2020A&A...640L...8U}; \citealt{2021ApJ...919....6K}). This trend is not sensitive to adopted $\alpha_{\rm CO}$ (Fig~\ref{fig:fgas}).
For the other three QGs which only have upper limits on molecular gas contents , ADF22-QG2/QG3 and GS-10578 (\citealt{2024arXiv240519401S}), such a scenario could also be applicable.

It is worth noticing that the AzTEC14 group is located within the cosmic web filaments and includes both DSFGs and QGs (Fig.~\ref{fig:alma_map}, \citealt{2019Sci...366...97U}). This might be weird: if abundant gas accretion were still occurring and fueling DSFGs in this group, it would challenge the passive nature of QGs, as the accreted gas could potentially reignite star formation in QGs. The observed situation can be explained if the cessation of cold gas accretion from the cosmic web filaments could occur at some point, possibly due to heating associated with virial shocks and/or starburst/AGN activity, and DSFGs are also on a track to a quiescent system (of various stages), consuming their existing gas reservoirs.

This scenario further aligns with the following observational clues: First, the fact that ADF22-QG1, QG2, and QG3 are closely located to each other and belong to a same galaxy group 
suggests a common mechanism driving their evolution. Second, Ly\,$\alpha$ emission from the filaments is relatively faint in this region (\citealt{2019Sci...366...97U}) in contrast with the high density of member galaxies (suggestive of plentiful amount of ionizing photons). This indicates limited availability of cool gas in the filaments for effective Ly\,$\alpha$ emission. Third, the large velocity dispersion within the group suggests the presence of a (nearly) collapsed massive halo with $M_{\rm vir}\sim10^{13.4-14}$\,M$_\odot$ (\citealt{2016MNRAS.455.3333K}), which hosts at least two X-ray AGNs (\citealt{2009MNRAS.400..299L}; \citeyear{2009ApJ...691..687L}) as possible heating sources. This massive halo may be sufficiently heated to suppress cold accretion.

Thus the gas starvation scenario likely plays a role in quenching, as highlighted by previous molecular gas studies of QGs (\citealt{2021ApJ...908...54W}; \citealt{2024arXiv240519401S}). However, the detection of a molecular gas reservoir in ADF22-QG1, despite its quiescent nature (Fig.~\ref{fig:gas_dust}b), suggests that star formation can be suppressed even in the presence of a significant amount of molecular gas. This suppression could be attributed to dynamical support preventing H$_2$ collapse. The stellar morphology of ADF22-QG1 is characterized by a combination of spheroidal and disk components (S\'ersic index $n=2.88\pm0.30$ at F444W, M.Kubo et al. in preparation), which makes morphological quenching (\citealt{2009ApJ...707..250M}) a plausible mechanism for maintaining its quiescent state.

\begin{deluxetable*}{lcccccccc}
\tabletypesize{\scriptsize}
\tablewidth{0pt} 
\tablecaption{Measured properties of ADF22-QGs \label{tab:measurement}}
\tablehead{
\colhead{ID} & \colhead{Coordinate} & \colhead{$z_{\rm spec}$} & \colhead{line}  & \colhead{log$M_*$} & \colhead{$S_{\rm CO}dV$} & \colhead{$M_{\rm H_2}^{\alpha4.4}$} & \colhead{$M_{\rm H_2}^{\alpha0.8}$} & \colhead{log$M_{\rm d}$} \\
\colhead{} & \colhead{ICRS} & \colhead{} & \colhead{} & \colhead{(M$_\odot$)} & \colhead{Jy\,m\,s$^{-1}$} & \colhead{(M$_\odot$)} & \colhead{(M$_\odot$)} }
\startdata 
ADF22-QG1 & 22:17:37.25, +00:18:16.0 & 3.0922$_{-0.004}^{+0.008}$ & Bal. abs. & 11.11 & 45 $\pm$ 8 & 10.26 $\pm$ 0.07  & 9.52 $\pm$ 0.07 & $<8.02$ \\
ADF22-QG2 & 22:17:37.29, +00:18:23.4 & 3.0851 $\pm$ 0.0001 & [O\,{\sc iii}], H$\beta$, [O\,{\sc ii}] & 10.74 & $<23$ & $<9.97$  & $<9.23$ & $<8.02$ \\
ADF22-QG3 & 22:17:37.12, +00:18:17.8 & 3.0744 $\pm$ 0.0003 & [O\,{\sc iii}] & 10.85 & $<23$ & $<9.97$  & $<9.23$ & $<8.02$  \\
\enddata
\tablecomments{
Spectroscopic redshifts ($z_{\rm spec}$) are adopted from literatures (\citealt{2015ApJ...799...38K}; \citeyear{2021ApJ...919....6K}; \citeyear{2022ApJ...935...89K}). 
Stellar masses ($M_*$) are calculated using FAST$++$ (e.g., \citealt{2021ApJ...919....6K}), updating with JWST NIRCam photometry. 
}
\end{deluxetable*}

\begin{acknowledgments}
We thank Ian Smail for useful discussions.
This work is based on observations made with the NASA/ESA/CSA James Webb Space Telescope. The data were obtained from the Mikulski Archive for Space Telescopes at the Space Telescope Science Institute, which is operated by the Association of Universities for Research in Astronomy, Inc., under NASA contract NAS 5-03127 for JWST. These observations are associated with program \#3547.
This paper makes use of the following ALMA data: ADS/JAO.ALMA\#2013.1.00162.S, \#2017.1.01332.S,\#2023.1.01206.S. ALMA is a partnership of ESO (representing its member states), NSF (USA) and NINS (Japan), together with NRC (Canada), NSTC and ASIAA (Taiwan), and KASI (Republic of Korea), in cooperation with the Republic of Chile. The Joint ALMA Observatory is operated by ESO, AUI/NRAO and NAOJ.
HU acknowledges support from JSPS KAKENHI Grant Numbers 20H01953, 22KK0231, 23K20240. 
%
%
This work was supported by NAOJ
ALMA Scientific Research Grant Numbers 2024-26A.
%

\end{acknowledgments}

\bibliography{sample631.bbl}{}
\bibliographystyle{aasjournal}



\end{document}